\documentclass{emulateapj}

\slugcomment{To appear in the Astrophysical Journal}

\shorttitle{Water in NGC~6334~I}
\shortauthors{Emprechtinger et al.}

\begin{document}

\title{The Abundance, Ortho/Para Ratio, and Deuteration of Water in the High-Mass Star Forming Region NGC~6334~I}

\author{M. Emprechtinger$^1$, D.~C.~Lis$^1$, R. Rolffs$^2$, P.~Schilke$^2$, R.~R.~Monje$^1$, C.~Comito$^3$, C.~Ceccarelli$^4$, D.~A.~Neufeld$^5$, and F.~F.~S. van der Tak$^6$}
\affil{$^1$California Institute of Technology, Cahill Center for Astronomy and Astrophysics 301-17, Pasadena, CA 91125, USA}

\affil{$^2$Physikalisches Institut, Universit\"at zu K\"oln, Z\"ulpicher Str. 77, 50937 K\"oln, Germany}

\affil{$^3$Max-Planck-Institut f\"ur Radioastronomie, Auf dem H\"ugel 69, 53121 Bonn, Germany}

\affil{$^4$UJF-Grenoble 1/ CNRS-INSU, Institut de Plan\'etologie et d'Astrophysique de Grenoble (IPAG) UMR 5274, Grenoble, F-38041, France}

\affil{$^5$Johns Hopkins University, Baltimore MD,  USA}

\affil{$^6$SRON Netherlands Institute for Space Research and Kapteyn Astronomical Institute, University of Groningen, Groningen, NL}

%\email{dcl@caltech.edu}

\begin{abstract} We present Herschel/HIFI observations of 30 transitions of water isotopologues toward the high-mass star forming region NGC~6334~I. The line profiles of H$_2^{16}$O, H$_2^{17}$O, H$_2^{18}$O, and HDO show a complex pattern of emission and absorption components associated with the embedded hot cores, a lower-density envelope, two outflow components, and several foreground clouds, some associated with the NGC~6334 complex, others seen in projection against the strong continuum background of the source.  Our analysis reveals an H$_2$O ortho/para ratio of $3 \pm 0.5$ in the foreground clouds, as well as the outflow. The water abundance varies from $\sim 10^{-8}$ in the foreground clouds and the outer envelope to $\sim 10^{-6}$ in the hot core. The hot core abundance is two orders of magnitude below the chemical model predictions for dense, warm gas, but within the range of values found in other Herschel/HIFI studies of hot cores and hot corinos. This may be related to the relatively low gas and dust temperature ($\sim 100$~K), or time dependent effects, resulting in a significant fraction of water molecules still locked up in dust grain mantles. The HDO/H$_2$O ratio in NGC~6334~I, $\sim 2 \times 10^{-4}$, is also relatively low, but within the range found in other high-mass star forming regions.  \end{abstract}


\section{Introduction}
Water is predicted to be one of the most abundant species in the interstellar medium (ISM) and, besides CO, the main reservoir of oxygen. In warm regions ($T_{dust} > 100$~K), where water sublimates from icy grain mantles, or is efficiently formed in the gas phase through reactions involving atomic oxygen,  its gas-phase abundance may be as high $10^{-4}$ with respect to H$_2$ (e.g., Ceccarelli et al. 1996; Doty et al. 2002; Rodgers \& Charnley 2003), although in the presence of strong UV radiation the water abundance is typically much lower, $\sim 10^{-7}$, due to photodissociation (Hollenbach et al. 2009). Consequently, water may be an important coolant, especially in high-mass star forming regions, where high gas and dust temperatures ($> 100$~K) and high densities ($> 10^6$~cm$^{-3}$) are common. At lower temperatures ($T_{dust} < 100$~K), water molecules freeze-out onto dust grains and the resulting gas-phase H$_2$O abundance is typically lower, by several orders of magnitude, than the chemical model predictions for cold clouds (e.g., Boonman et al. 2003; Doty et al. 2002). The freeze-out of water in cold clouds has been confirmed by the detection of water ice (e.g., Willner et al. 1982; Keane et al. 2001; Gibb \& Whittet 2002). Because of the potentially important role of water in the energy balance in star forming regions, and therefore its influence on the collapse and fragmentation of cloud cores, its abundance is an important parameter in understanding the star formation process. However, because of the complex interplay between different routes of the gas chemistry, grain-surface chemistry, freeze-out and desorption mechanisms, as well observational challenges, the chemistry of water in the interstellar medium (ISM) is not yet well understood.

Most low-excitation H$_2$O lines, especially the ground state ortho and para transitions, are not observable from the ground, because of the low atmospheric transmission due to the terrestrial water vapor. Space missions have already provided ample water observations in various ISM environments (van Dishoeck \& Helmich 1996---ISO; Melnick \& Bergin 2005---SWAS; Bjerkeli et al. 2009---Odin). However, these have been limited to a small number of transitions (SWAS, Odin), low spectral resolution (ISO), and low spatial resolution ($> 1^\prime$). Previous observations of high-mass star forming regions indicate an H$_2$O abundance as high as $\sim 10^{-4}$ in warm ($T_{dust} > 100$~K) regions, whereas in dense environments at temperatures below 100~K, where freeze-out takes place, the H$_2$O abundance can be as low as $10^{-8}$ (e.g., Boonman et al. 2003; Snell et al. 2000; Bergin \& Snell 2002; Herpin et al. 2012). The water abundance in low-mass star forming regions ranges from $5 \times 10^{-7}$ in the envelopes to $\sim 10^{-6}$ in the hot corinos (Ceccarelli et al. 2000) and $10^{-5}$ in the outflows of Herbig Haro objects (Liseau et al. 1996). In diffuse clouds water abundances between $10^{-7}$ and $10^{-6}$ have been derived (Cernicharo et al. 1997; Moneti et al. 2001).

Here we present water observations carried out with the HIFI instrument (de Graauw et al. 2010) aboard the Herschel Space Observatory (Pilbratt et al. 2010), which has provided for the first time the ability to observe multiple water lines with high spectral and spatial resolution.  Using this instrument, several investigations of water in star forming regions have already been carried out. Melnick et al. (2010) derived a water abundance of $1 - 7 \times 10^{-5}$ in Orion KL, in agreement with chemical model predictions for warm gas. The Herschel guaranteed time key program ``Water In Star-Forming regions with Herschel'' (WISH; Van Dishoeck et al. 2011) is dedicated to investigations of water in some 80 galactic sources. Early WISH results show that the H$_2$O abundance in the shocked gas in NGC~1333 is similar to the abundance in Orion KL (Kristensen et al. 2010), whereas Caselli et al. (2012) derived a much lower water abundance of $\geq10^{-9}$ within the central few thousand astronomical units in the prestellar core L1544.  Chavarr\'{\i}a et al. (2010) investigated water in the high-mass protostar W3~IRS5 and derived an H$_2^{16}$O abundance of $10^{-4}$ in the hot core and $10^{-9} - 10^{-8}$ in the cooler envelope. All these measurements can be explained by theoretical models. However, van der Tak et al. (2010) derived an H$_2$O abundance of only $2 \times 10^{-10}$ in the dense core and $4 \times 10^{-9}$ in the cooler envelope of the high-mass starforming region DR~21.  Flagey et al. (2013) derived a water abundance of $5\times 10^{-8}$ with respect to H$_2$ in the diffuse clouds in the Galactic disk, while Sonnentrucker et al. (2013) found a factor of 3 enhancement in the water abundance in the material within 200~pc from the Galactic center, as compared to the disk.  These strong variations in the derived water abundance demonstrate the importance of fully understanding the water chemistry in a wide variety of interstellar environments.

The abundance of HDO and the D/H ratio in water is another key parameter, because it is an indicator of the formation temperature. From the cosmic D/H ratio ($1.5 \times 10^{-5}$; Oliveira et al. 2003) one would expect an HDO/H$_2$O ratio of $3 \times 10^{-5}$. However, strong enhancements in abundances of deuterated molecules, due to their lower zero-point vibrational energies as compared to the normal variants (Phillips \& Vastel 2003), take place at temperatures $< 50$~K (Roberts \& Millar 2000). Since water is largely frozen-out at these low temperatures, ices on dust grains contain a high fraction (0.2\%--2\%) of deuterated water (Dartois et al. 2003; Parise et al. 2003). HDO has been observed in Sgr B2(M) and (N) (Comito et al. 2003, 2010) with abundances varying from $2.5 \times 10^{-11}$ ($T_{gas} < 100$~K) to $3.5 \times 10^{-9}$ ($T_{gas} > 200$~K)\footnote{At the molecular hydrogen densities typical of dense interstellar clouds, the gas and dust are usually well coupled, $T_{gas}=T_{dust}$.}, and HDO/H$_2$O$ \sim 10^{-3}$. In the low-mass protostar L1448-mm a very high HDO/H$_2$O ratio of 0.04 was found (Codella et al. 2010), which agrees with the abundance ratios in other low-mass protostars such as IRAS 16293-2422 (0.2\%--3\%; Parise et al. 2005) and NGC~1333 IRAS2 ($> 1$\%; Liu et al. 2011). Observations of sites of high-mass star formation, which show in general lower abundances of deuterated molecules, revealed HDO/H$_2$O ratios between $1 \times 10^{-4}$ (W3 IRS5; van der Tak et al. 2006) and 0.02 (Orion KL; Bergin et al. 2010). The high abundance of HDO in these sources indicates that water molecules have been predominately formed at a cold, earlier evolutionary stage and have just recently been evaporated from dust grains into the gas phase, where they can be studied by means of submillimeter spectroscopy.

In this paper, we analyze spectra of water isotopologues toward NGC~6334~I, a relatively nearby (1.7 kpc; Neckel 1978) high-mass star forming region, which harbors sites of many stages of protostellar evolution (Straw \& Hyland 1989). Single-dish submillimeter continuum observations indicate a total gas mass of 200~M$_\sun$ associated with NGC~6334~I (Sandell 2000). The molecular hot core emission has been studied extensively over the past decade (e.g., Beuther et al. 2008, 2007; Hunter et al. 2006). The source is associated with an ultra compact HII region (de Pree et al. 1995) and shows a very line-rich spectrum (Schilke et al. 2006; Thorwirth et al. 2003). Furthermore, a bipolar outflow (Leurini et al. 2006; Beuther et al. 2008) and H$_2$O, OH, CH$_3$OH class II, and NH$_3$ masers have been detected (e.g., Kraemer \& Jackson 1995; Ellingsen et al. 1996; Walsh et al. 2007). Rolffs et al. (2011) studied the radial density structure of several high-mass star forming cores, including NGC~6334~I, mainly based on single-dish HCN observations with the APEX telescope, and approximated the source structure as a centrally heated sphere with a power-law density distribution. In the case of NGC~6334~I, this leads to a density law $n \sim r^{-1.5}$, which describes well the envelope traced by the single-dish observations. However, interferometric data (SMA) reveal that the internal structure of NGC~6334~I is much more complex---the hot core consists of four compact condensations within a $10^{\prime\prime}$ region that contain about 50\% of the single-dish continuum flux (Hunter et al. 2006). High-resolution molecular observations, combined with 3-d radiative transfer models, are required to fully understand the distribution of water in NGC~6334~I, however, simple models like the ones presented here already provide a good starting point for a more sophisticated analysis of the HIFI data.

The data presented here are taken from an unbiased HIFI spectral line survey, carried out as part of the Herschel guaranteed time key program ``Chemical HErschel Surveys of Star forming regions'' (CHESS; Ceccarelli et al. 2010). In an earlier paper (Emprechtinger et al. 2010, hereafter paper I), we presented a preliminary analysis of twelve water lines in NGC~6334~I, including the $1_{11} - 0_{00}$ and $1_{10} - 1_{01}$ fundamental rotational transitions, observed as a part of the Herschel science demonstration program (SDP). Assuming that all ortho and para molecules are in their respective ground states, we determined an ortho/para ratio of $\sim 1.6 \pm 1$ (3$\sigma$ uncertainty) in the foreground clouds, which appeared lower than the statistical value of 3, as seen for example in the outflow. The water abundance ranges between $10^{-8}$ and $4 \times 10^{-5}$, in agreement with theoretical models and previous findings (e.g. Rodgers \& Charnley 2003; Doty et al.  2002; Ceccarelli et al. 1996; Herpin et al. 2012). In addition, the H$_2^{18}$O/H$_2^{17}$O ratio of $3.7 \pm 0.6$ agrees well with the isotopic $^{18}$O/$^{17}$O ratio derived from formaldehyde measurements (Wilson \& Rood 1994). 

Here we present the complete HIFI data set of water observations toward NGC~6334~I, including 30 lines of H$_2^{16}$O, H$_2^{18}$O, H$_2^{17}$O, and HDO. In the following sections, we present the observations (\S 2), along with a simple LTE analysis (\S 3), and full radiative transfer modeling of H$_2^{16}$O lines (\S~4). We summarize the results in \S 5.

\newpage
\section{Observations}\label{obs}
HIFI observations of NGC~6334~I ($\alpha_{J2000}$ = 17$^{\rm h}$20$^{\rm m}$53.32$^{\rm s}$, $\delta_{J2000}$ = $−35^\circ$46$^\prime$58.5$^{\prime\prime}$) presented here were carried out between 2010 February 28 and October 14, in the double beam switch mode (180$^{\prime\prime}$ chopper throw). The data have been reduced using the Herschel Interactive Processing Environment (HIPE; Ott 2010) version 5.1. The spectral resolution of the double sideband (DSB) spectra, obtained using the HIFI WBS spectrometer, is $\sim 1.1$~MHz. The DSB spectra have been observed with a redundancy of 8 for the SIS mixer bands (8 LO settings per IF bandwidth) and 4 for the HEB bands, which allows for the deconvolution and isolation of the single sideband (SSB) spectra (Comito \& Schilke 2002). The deconvolved SSB spectra have been exported to the FITS format for subsequent analysis using the IRAM GILDAS package\footnote{http://www.iram.fr/IRAMFR/GILDAS/}. Spectra presented here are equally weighted averages of the H and V polarizations, corrected for the corresponding aperture efficiencies (Roelfsema et al. 2012; Table~1).

In the CHESS spectral survey, we detect sixteen H$_2^{16}$O, six H$_2^{18}$O, four H$_2^{17}$O, and four HDO lines.
Several other lines, especially of the rarer isotopologues, are blended with emission lines of other molecules, and thus provide only upper limits for the analysis. In Table~1 we summarize the observing parameters of the detected water lines.

\begin{table*}
\begin{center}
\caption{Water transitions observed in NGC~6334~I.\label{linepara}}
\begin{tabular}{llccccc}
\hline\hline
$Isotopologue$ & $Transition$ & $HIFI\ Band$ & $Frequency$ & $E_{up}$ & $Beam\ Size$ & $\eta _{ap}$\\
& & & (GHz) & (K) & ($''$) &\\
\hline
 p-H$_2^{16}$O & $6_{24}-7_{17}~^*$ & 1A & 488.491 & 867.6 & 43.4 & 0.68\\
& $2_{11}-2_{02}~^*$ & 2B & 752.033 & 137.0 & 28.2 & 0.67\\
& $4_{22}-3_{31}~^*$ & 3B & 916.172 & 454.5 & 23.2 & 0.67\\
& $5_{24}-4_{31}~^*$ & 4A & 970.315 & 599.1 & 21.9 & 0.65\\
& $2_{02}-1_{11}$ & 4A & 987.927 & 100.9 & 21.5 & 0.65\\
& $1_{11}-0_{00}~^*$ & 4B & 1113.343 & 53.5 & 19.0 & 0.65\\
& $4_{22}-4_{13}$ & 5A & 1207.639 & 454.5 &17.6 & 0.55\\
& $2_{20}-2_{11}$ & 5A & 1228.789 & 196.0 &17.3 & 0.55\\
\hline
o-H$_2^{16}$O & $1_{10}-1_{01}~^*$ & 1A & 556.936 & 61.0 & 38.1 & 0.68\\
& $5_{32}-4_{41}$ & 1B & 620.701 & 732.4 &34.2 & 0.67\\
& $3_{12}-3_{03}$ & 4B & 1097.365 & 249.5 &19.3 & 0.65\\
& $3_{12}-2_{21}~^*$ & 5A & 1153.127 & 249.5 & 18.4 & 0.55\\
& $6_{34}-5_{41}$ & 5A & 1158.323 & 934.2 &18.3 & 0.55\\
& $3_{21}-3_{12}$ & 5A & 1162.912 & 305.4 & 18.2 & 0.55\\
& $2_{21}-2_{12}$ & 6B & 1661.008 & 194.2 & 12.8& 0.63 \\
& $2_{12}-1_{01}$ & 6B & 1669.905 & 114.4 & 12.7  & 0.63 \\
\hline
p-H$_2^{18}$O & $2_{11}-2_{02}~^*$ & 2B & 745.320 & 136.4 & 28.4 & 0.67\\
& $2_{02}-1_{11}$ & 4A & 994.675 & 100.7 & 21.3 & 0.65\\
& $1_{11}-0_{00}~^*$ & 4B & 1101.698 & 52.9 &19.3 & 0.65\\
\hline
o-H$_2^{18}$O & $1_{10}-1_{01}$ & 1A & 547.676 & 60.5 & 38.8 & 0.68\\
& $3_{12}-3_{03}$ & 4B & 1095.627 & 248.9 & 19.4 & 0.65\\
& $2_{12}-1_{01}$ & 6B & 1655.867 & 113.7 & 12.8 & 0.63\\
\hline
p-H$_2^{17}$O & $2_{11}-2_{02}$ & 2B & 748.458 & 136.7 & 28.3 & 0.67\\
& $1_{11}-0_{00}~^*$ & 4B & 1107.167 & 53.2 &19.2 & 0.65\\
\hline
o-H$_2^{17}$O & $1_{10}-1_{01}~^*$ & 1A & 552.020 & 60.7 &38.4 & 0.68\\
& $3_{12}-3_{03}$ & 4B & 1096.414 & 249.2 & 19.3 & 0.65\\
\hline
HDO & $2_{11}-2_{02}$ & 1B & 599.927 & 95.3 & 35.3 & 0.67\\ 
& $1_{11}-0_{00}$ & 3B & 893.639 & 43.9 & 23.7 & 0.67\\
& $2_{02}-1_{01}$ & 3B & 919.311 & 66.5 & 23.1 & 0.67\\
& $2_{11}-1_{10}$ & 4A & 1009.945 & 95.3 & 21.0 & 0.65\\
\hline
\end{tabular}
\end{center}
Notes---Entries in the table are: the isotopologue and transition, the HIFI mixer band, the rest frequency (GHz), the upper state energy (K), the Herschel beam size (arcsec), and the aperture efficiency (Roelfsema et al. 2012). An asterisk marks lines that have already been presented in paper~I.
\end{table*}  

\section{Results \& Analysis}\label{RaA}

\begin{figure*}
\begin{center}
\includegraphics[angle=90,scale=0.95]{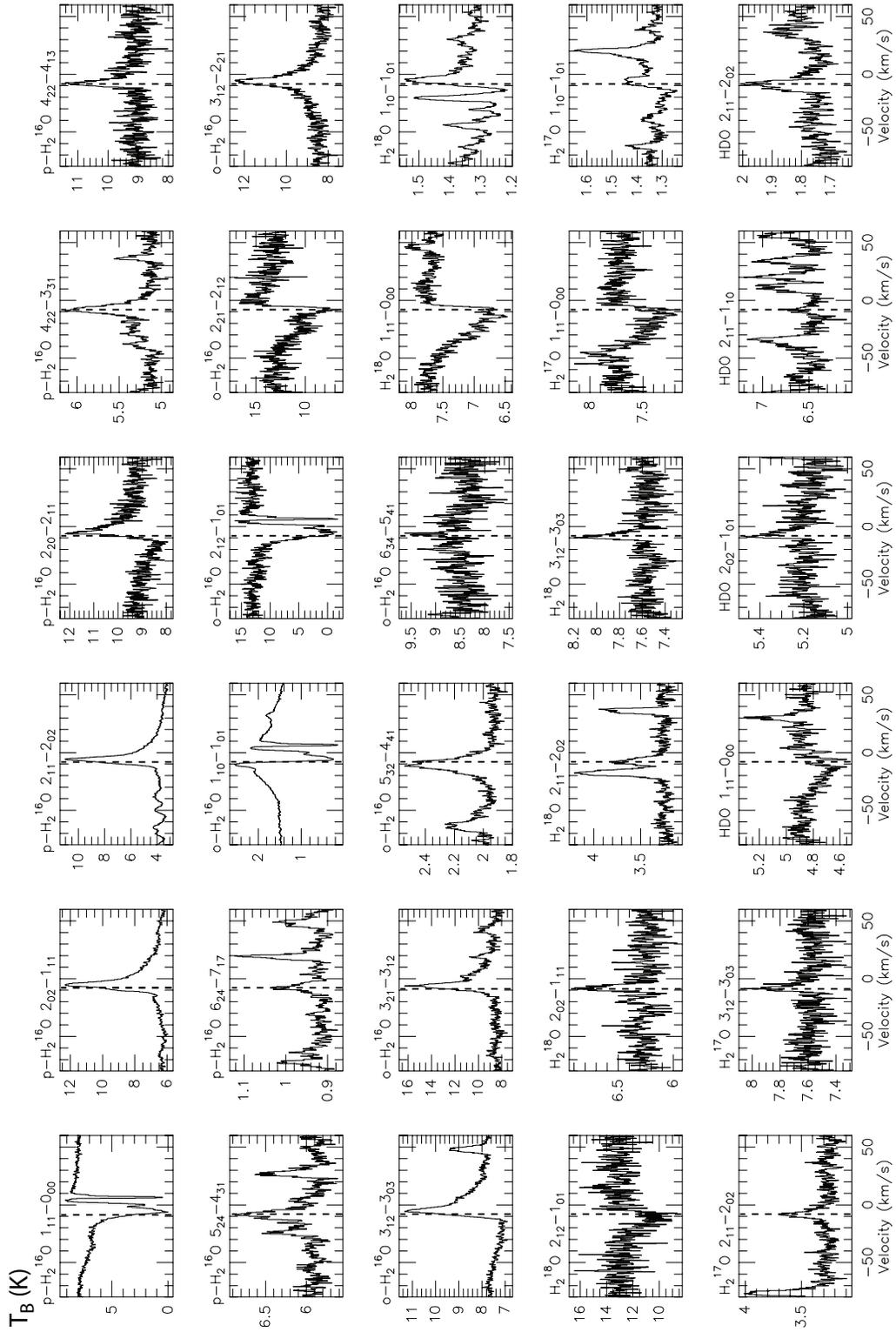}
\caption{HIFI spectra of the 30 transitions of H$_2^{16}$O, H$_2^{17}$O, H$_2^{18}$O, and HDO toward NGC~6334~I. The intensity scale is brightness temperature, in K, corrected for the aperture efficiency. Dashed vertical lines mark the velocity of the dense core SMA~2 at $-8$~km\,s$^{-1}$.}\label{allspec}
\end{center}
\end{figure*}

\begin{table*}
\begin{center}
\caption{Physical properties of the gas components in NGC~6334~I.\label{components}}
\begin{tabular}{lcccccccc}
\hline\hline
  & $Envelope$ & $Hot\ Core$ & \multicolumn{2}{c}{$Outflow$} & \multicolumn{4}{c}{$Foreground$} \\
  & & & $Broad$ & $Narrow$ & $1$ & $2$ & $3$ & $4$\\  
\hline
$v_{\rm LSR}$ (km\,s$^{-1}$)           & --6.4 & --8.2 & --10 & --10 & --3 &    0 &+6.5 & +8\\
$\Delta v$ FWHM (km\,s$^{-1}$) &     5.4 &     4.0 &      80 &     32 & 5.0  & 3.0 & 1.2 & 2.5 \\
$T_{gas}$ (K)                                  & 20--100  & $>100$ & $>50$ & 130 & \multicolumn{4}{c}{$\sim 30$} \\ 
$n$(H$_2$) (cm$^{-3}$)                & $5 \times 10^4$--$5 \times 10^6$ & $10^7$--$10^8$ & -- & $2\times 10^7$ & \multicolumn{4}{c}{$<10^4$}\\
\hline
\end{tabular}
\end{center}
\end{table*}

\subsection{Line Shapes}

Spectra of all 30 water transitions observed in NGC~6334~I are displayed in Figure~1. Their line shapes differ significantly, indicating that different lines trace different regions within the beam. From the line shapes, we can identify four major components. The first component, seen in many transitions, has a Gaussian line shape at a velocity of approximately --6.4 km\,s$^{-1}$, with a line width of approximately 5 km\,s$^{-1}$. This feature corresponds to the envelope of NGC~6334~I. In the ground state transitions of H$_2^{16}$O, but also in the H$_2^{16}$O $2_{21} - 2_{12}$ line, this component is seen in absorption against the background continuum; in most other transitions it is seen in emission. The second component is identified with one of the embedded cores. Some emission lines, especially those of highly excited lines of H$_2^{16}$O and lines of rare isotopologues, have central velocities close to --8 km\,s$^{-1}$, corresponding to the local standard of rest (LSR) velocity of the embedded core SMA~2 (Beuther et al. 2007). Thus we associate these lines with the dense, compact core.

The third major component are the foreground clouds, some of which reside in the immediate vicinity of NGC~6334~I, while others are not physically associated with the NGC~6334 complex. van der Wiel et al. (2010) identified four foreground components at --3, 0, +6.5, and +8 km\,s$^{-1}$ based on HIFI observations of CH. Three of these components (0, +6.5, and +8 km\,s$^{-1}$) are also detected in the ground state H$_2^{16}$O transitions. Whereas the 0 km\,s$^{-1}$ component most likely originates in the local environment of NGC~6334~I, the features at +6.5 and +8 km\,s$^{-1 }$ are associated with clouds in the Sagittarius spiral arm.

The fourth major spectral component are the outflows, evidence of which is seen in almost all H$_2^{16}$O lines with $J_u \leq 3$, as well as all ground-state transitions of the rarer isotopologues.  In some transitions the outflows are detected in emission, in others in absorption, and in some cases P-Cygni profiles are seen.

The physical properties of these four physical components are summarized in Table~2.  In the following sections we investigate in detail three of these components: the dense core, the outflow, and the foreground clouds, individually. Table~3 lists all transitions observed and specifies which components they trace.

\begin{table*}
\begin{center}
\caption{Spectral components seen in the water spectra.}{\label{linshape}}
\begin{tabular}{llccccc}
\hline\hline
$Isotopologue$ & $Transition$ & $Frequency$ & $Main$ & $Dense$ & $Outflow$ & $Foreground$ \\
& & (GHz) & $Component$ & $Core$ & & \\
\hline
 p-H$_2^{16}$O & $6_{24}-7_{17}$ & 448.5 & N & Y & N & N \\
& $2_{11}-2_{02}$ & 752.0 & E & N& Y & N \\
& $4_{22}-3_{31}$ & 916.2 & N & Y & N & N \\
& $5_{24}-4_{31}$ & 970.3 & N & Y & N & N \\
& $2_{02}-1_{11}$ & 987.9 & E & N & Y & N \\
& $1_{11}-0_{00}$ & 1113.3 & A & N &P$^b$ & Y \\
& $4_{22}-4_{13}$ & 1207.6 & N & Y & N & N \\
& $2_{20}-2_{11}$ & 1228.8 & E & N & P & N \\
\hline
o-H$_2^{16}$O & $1_{10}-1_{01}$ & 556.9 & A & N & Y$^b$ & Y \\
& $5_{32}-4_{41}$ & 620.7 & N & Y & N & N \\
& $3_{12}-3_{03}$ & 1097.4 & E & N & P & N \\
& $3_{12}-2_{21}$ & 1153.1 & E & N & Y & N \\
& $6_{34}-5_{41}$ & 1158.1 & N & Y & N & N \\
& $3_{21}-3_{12}$ & 1162.9 & E & N & N & N \\
& $2_{21}-2_{12}$ & 1661.0 & A & N & P & N \\
& $2_{12}-1_{01}$ & 1669.9 & A & N & P$^b$ & Y \\
\hline
p-H$_2^{18}$O & $2_{11}-2_{02}$ & 745.3 & N & Y & N & N \\
& $2_{02}-1_{11}$ & 994.7 & N & Y & N & N \\
& $1_{11}-0_{00}$ & 1101.7 & N & N & A$^b$ & N \\
\hline
o-H$_2^{18}$O & $1_{10}-1_{01}$ & 547.7 & N & N & A$^b$  & N\\
& $3_{12}-3_{03}$ & 1095.6 & N & Y & N & N \\
& $2_{12}-1_{01}$ & 1655.9 & N & N & A$^b$ & N\\
\hline
p-H$_2^{17}$O & $2_{11}-2_{02}$ & 748.5 & N & Y & N & N \\
& $1_{11}-0_{00}$ & 1107.2 & N & N & A$^b$ & N \\
\hline
o-H$_2^{17}$O & $1_{10}-1_{01}$ & 552.0 & N & N & A$^b$ & N \\
& $3_{12}-3_{03}$ & 1096.4 & N & Y & N & N \\
\hline
HDO & $2_{11}-2_{02}$ & 599.9 & N & Y & N & N \\ 
& $1_{11}-0_{00}$ & 893.6 & N & N & A$^b$ & N \\
& $2_{02}-1_{01}$ & 919.3 & N & Y& N & N \\
& $2_{11}-1_{10}$ & 1009.9 & N & Y & N & N \\
\hline
\end{tabular}
\end{center}
Notes---{\it{Main Component:}} E= seen in emission, A=seen in absorption. \emph{Dense Core}: Y=a component detected at a velocity of $-8$~km\,s$^{-1}$, and thus associated with the dense cores. {\it{Outflow:}} Y=outflow detected in emission, A=blue lobe of the outflow in absorption, P= P-Cygni profile. The superscript $b$ indicates that the broad outflow component is seen. {\it{Foreground:}} Y=foreground components are detected. N in all three columns means that this component does not show up in the respective transition.
\end{table*}

\subsection{Foreground Clouds}

\begin{figure}
\begin{center}
\includegraphics[angle=0,width=0.7\columnwidth]{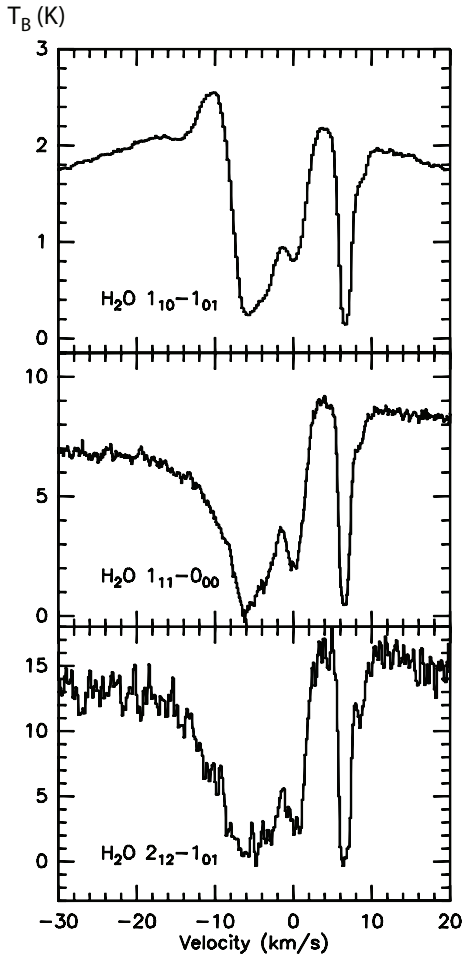} 
\caption{Spectra of the three ground state transitions of ortho- and para-H$_2^{16}$O toward NGC~6334~I. The intensity scale is brightness temperature, in K, corrected for the aperture efficiency.}\label{fgfig} 
\end{center}
\end{figure}

\subsubsection{Ortho/Para Ratio Revisited}

Foreground absorption features are best seen in the ground state transitions of para-H$_2^{16}$O ($1_{11} - 0_{00}$) and ortho-H$_2^{16}$O ($1_{10} - 1_{01}$, $2_{12} -1_{01}$ ), which are displayed in Figure~2. In paper I, we determined the ortho/para ratio in the foreground components using the $1_{10} - 1_{01}$ and the $1_{11} - 0_{00}$ lines. Under the assumptions that all ortho and para water molecules are in their respective ground states, these observations yield a low ortho/para ratio of $1.6 \pm 1$ (3$\sigma$ uncertainty), below the statistical value of 3. With the additional observations of the second ground state ortho-water line, the $2_{12} - 1_{01}$ transition at 1669~GHz, we can independently constrain the excitation temperature, assuming a level population according to the Boltzmann distribution and determine the ortho/para ratio more reliably.  In order to derive the ortho/para ratio, we first resample the spectra to a common velocity resolution of 0.27~km\,s$^{-1}$, which allows us to compare column densities in individual velocity bins.  The optical depth of the ground state transitions is calculated as
\begin{equation}
\tau = -ln\left(\frac{T_{B}-J_\nu(T_{ex})}{T_{back}-J_\nu(T_{ex})}\right)
\label{tau}
\end{equation}
where $T_{B}$ is the measured brightness temperature, $T_{back}$ is the corresponding background temperature, and $J_\nu(T_{ex})=\frac{h\nu}{k}/(e^{h\nu/kT_{ex}}-1)$ is the Rayleigh-Jeans equivalent radiation field due to the molecular emission. We neglect the influence of the cosmic microwave background (CMB), which is negligible at the frequencies considered here ($< 1.5$~mK).
In the case of the $1_{11}-0_{00}$ and $2_{12}-1_{01}$ lines, the background radiation is well defined by the continuum level at nearby frequencies. For the $1_{10}-1_{01}$ line, however, the background is dominated by the high-velocity emission from the outflow. We thus applied a linear baseline fit in the vicinity of the absorption components in order to determine the correct background level for the optical depth computations.\footnote{We assume that the absorption completely covers the continuum. This is a reasonable assumption, as for example the OH data of Brooks \& Whiteoak (2001) show that the 6 km\,s$^{-1}$ absorption is seen not only toward NGC~6334, but also toward other nearby HII regions. In addition, the deepest absorption seem in the ground state water lines is very close to zero.} Optical depths less than 2.5 are determined quite reliably, however, for $\tau>2.5$, the uncertainty of $\tau$ is comparable to $\tau$. Thus we do not include these data points in the subsequent analysis.  From the optical depth, we determine the population in the ground state per velocity interval as
\begin{equation}
N_{ground}=\frac{8\pi}{\lambda^3 A}\frac{g_l}{g_u}\frac{\tau}{1-e^{-h\nu/kT_{ex}}}
\label{Ntau}
\end{equation} 
where $\lambda$ is the wavelength, $A$ the Einstein spontaneous emission coefficient, $g_l$ and $g_u$ the statistical weights for the lower and the upper state, respectively, and $T_{ex}$ the excitation temperature.

Combining equations (1) and (2), we obtain for each transition an equation with two unknowns, $N_{ground}$ and $T_{ex}$. Assuming a level population according to the Boltzmann distribution implies the same excitation temperature for all three transitions detected in the foreground clouds. Furthermore, the population of the ground state for the two ortho-H$_2$O lines is identical, as they both connect to  the $1_{01}$ level. We can thus eliminate it from the equations and compute the excitation temperature directly. This leads to $T_{ex} =6.5$~K.

As a consistency check, we derive an upper limit for the population in the $1_{11}$ para-H$_2$O state, and thus for the excitation temperature, from the non-detection of the foreground clouds in the $2_{02} - 1_{11}$ transition. The upper limit on the optical depth in this line is 0.009 leading to an upper limit of $T_{ex} \leq 8.7$~K, in agreement with the value derived from the ortho-H$_2$O lines.

Equations (1) and (2), with the derived excitation temperature of 6.5 K, lead to an ortho/para ratio of $3 \pm 0.5$ in the foreground clouds, in good agreement with the statistical value of 3. Figure~3 shows the calculated ortho/para ratio as a function of velocity.  Comparing these results with the simple analysis in paper I (assuming that all population is in the respective ground state, leading to an ortho/para ratio of 1.6) reveals that the influence of the excitation temperature on the derived column density of the para-H$_2$O is negligible, due to the high energy of the first excited state ($1_{11}$, $E$=53.4~K). Even the partition function of ortho-H$_2$O changes by less than 3\%. The main reason for the discrepancy between the two methods is that the quantity $J_\nu(T_{ex})$ at the frequency of the $1_{10}-1_{01}$ ortho-H$_2$O line, 557~GHz, is a sensitive function of $T_{ex}$. Thus the optical depth and therefore the column density of ortho-H$_2$O was underestimated in paper I by approximately a factor of 1.9, leading to the low value of 1.6 for the ortho/para ratio.

\begin{figure}
\begin{center}
\includegraphics[width=0.9\columnwidth]{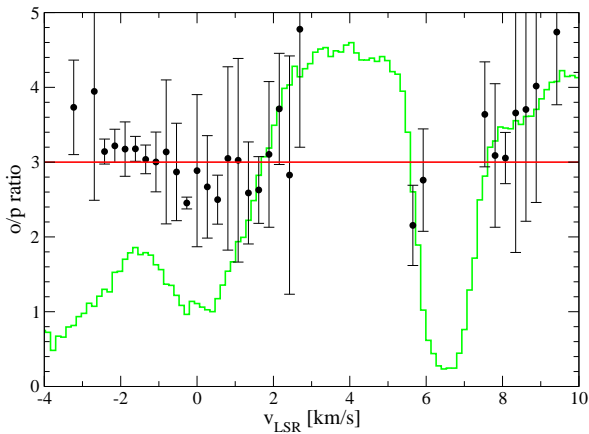}
\caption{Water ortho/para ratio toward NGC~6334~I in the velocity range from --4 to 10 km\,s$^{-1}$, corresponding to the foreground clouds. Green line shows the p-H$_2$O absorption (middle panel in Fig.~2). Red line indicates the statistical limit of 3.} \label{opres}
\end{center}
\end{figure}

The assumption of a thermalized level population may not be fully justified, considering the high critical densities of $6.4 \times 10^7$ cm$^{-3}$ and $1.12 \times 10^8$ cm$^{-3}$ of the $1_{10} - 1_{01}$ and $2_{12} - 1_{01}$ transitions, respectively. However, since the critical density of the $2_{12} - 1_{01}$ line is higher, it is the excitation temperature of this transition, which would be overestimated by assuming LTE. But, as shown above, only the $1_{10} - 1_{01}$ line is sensitive to changes in $T_{ex}$. Therefore a deviation from the Boltzmann distribution does not have significant influence on the results presented here.

An excitation temperature of 6.5 K for the H$_2$O is relatively high for diffuse clouds. However, some of the foreground components toward NGC~6334~I are also seen in absorption in the ground state para-NH$_3$ transitions, which indicates the presence of gas denser ($\sim 10^5$ cm$^{-3}$) than typically found in diffuse clouds. This analysis nicely illustrates the importance of the HIFI capabilities to observe multiple transitions for the determination of accurate column densities and molecular abundances.

\subsubsection{Water Abundance in the Foreground Clouds}

To determine the water abundance in the foreground clouds, we use the HF (1--0) absorption spectrum (Emprechtinger et al. 2012)  to estimate the H$_2$ column density. HF has been established as an excellent tracer of H$_2$ (Neufeld et al. 2005; Neufeld et al. 2010; Sonnentrucker et al. 2010; Monje et al. 2011).  Following these studies, we use an HF abundance of $1 \times 10^{-8}$ with respect to H$_2$, as determined on several lines of sight. Because of the high Einstein A coefficient ($2.423 \times 10^{−2}$ s$^{-1}$) and the low collisional excitation rates (Guillon et al. 2011), the critical density of HF (1--0) is $\sim 6 \times 10^9$ cm$^{-3}$, and thus approximately three orders of magnitude higher than the critical density of the ortho-H$_2$O $1_{10} - 1_{01}$ transition. Therefore, we assume that all HF molecules are in the ground state (see Emprechtinger et al. 2012). Unfortunately, the HF (1--0) line, like H$_2^{16}$O , quickly becomes optically thick ($\tau > 2.5$) at the line center and therefore we can determine the HF column density only in the line wings, where we derive an average H$_2$O abundance of $1 - 3 \times 10^{-8}$ in the foreground clouds (see Fig.~4), comparable to typical values determined in low-temperature regions ($T_{dust} < 100$~K).

\begin{figure}
\begin{center}
\includegraphics[angle=-90,width=0.9\columnwidth]{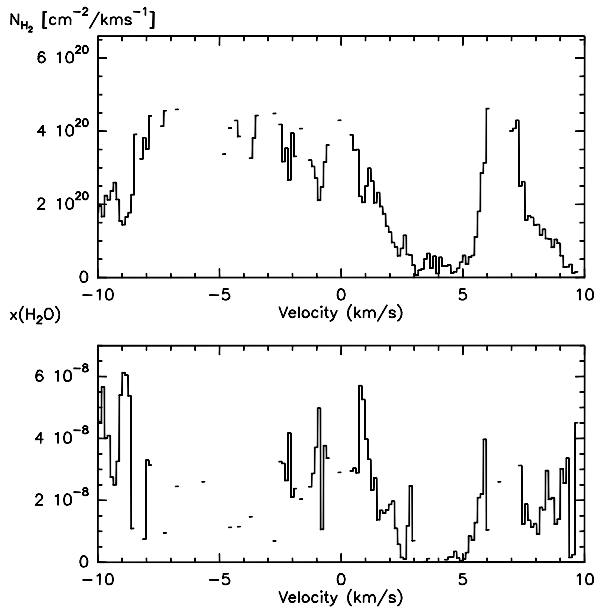}
\caption{H$_2$ column density (upper panel, based on HF absorption data) and H$_2$O abundance (lower panel) in the foreground clouds on the line of sight toward NGC~6334~I, plotted as a function of the LSR velocity.} \label{abufig}
\end{center}
\end{figure}

\subsection{Outflows}

In paper I, we have already shown that NGC~6334 I harbors two outflow components. One of these components shows broad lines (from --85 to +65 km\,s$^{-1}$ ) similar to the CO outflow detected by Leurini et al. (2006), who derived a gas temperature $> 50$~K and a density $> 1 \times 10^4$ cm$^{-3}$ at the peak position of the red lobe, and $T_{gas}> 15$~K and $n> 1 \times 10^3$~cm$^{-3}$ toward the blue lobe. This outflow component was already analyzed in paper I, where we determined that most of the water molecules are in the ground state, as this outflow component is not seen in any excited lines.  This is in contrast to HIFI observations of other high-mass star forming regions that do show bright high-velocity emission in excited water lines (e.g. W3 IRS5; Chavarr\'{\i}a et al. 2010) and may be related to the relatively low density in the outflowing gas (Leurini et al. 2006), resulting in subthermal excitation of water molecules.  Based on the H$_2^{17}$O and H$_2^{18}$O observations, we derived and optical depth of $\sim 300$ for the H$_2$O $1_{11} - 0_{00}$ transition in the velocity range --35 to --20 km\,s$^{-1}$. A comparison with LVG models (Radex; van der Tak et al. 2007) shows that our findings are in agreement with the lower limits for the temperature and density found by Leurini et al. (2006). However, despite its high optical depth, this outflow component does not absorb all background radiation, leading to the conclusion that only approximately 17\% of the total continuum of NGC~6334~I is subject to absorption by the outflow (area coverage factor of 17\%). The water abundance in the broad line outflow is $4 \times 10^{-5}$ (paper I), due to sublimation or sputtering of water ices. The second outflow component is much narrower, ranging from --45 to +25 km\,s$^{-1}$. Furthermore, the narrow outflow is also seen in some excited lines, which indicates that this component is warmer and denser. It is not clear, if these two outflow components represent two separate outflows or different regions within one outflow. Within the measurement uncertainties, the central velocity of both outflow components is --10 km\,s$^{-1}$, which is consistent with the latter scenario. This is further supported by interferometric observations of Beuther et al. (2008) who found only one outflow in their HCN (1--0) data.

\subsubsection{Warm and Dense Outflow}\label{NOut}

The three ground state water transitions are not very well suited to investigate the warm and dense outflow component, because the much more massive, but less excited broad outflow dominates the spectra. Furthermore, foreground absorption features hamper a clear view of the high-density outflow. However, the dense outflow features can be seen clearly in the moderately excited $2_{02} - 1_{11}$, $2_{11} - 2_{02}$, and $2_{20} - 2_{11}$ para-H$_2$O lines, as well as the $3_{12} - 3_{03}$ ortho-H$_2$O line (Figure 5).

\begin{figure*}
\begin{center}
\includegraphics[angle=0,width=0.7\textwidth]{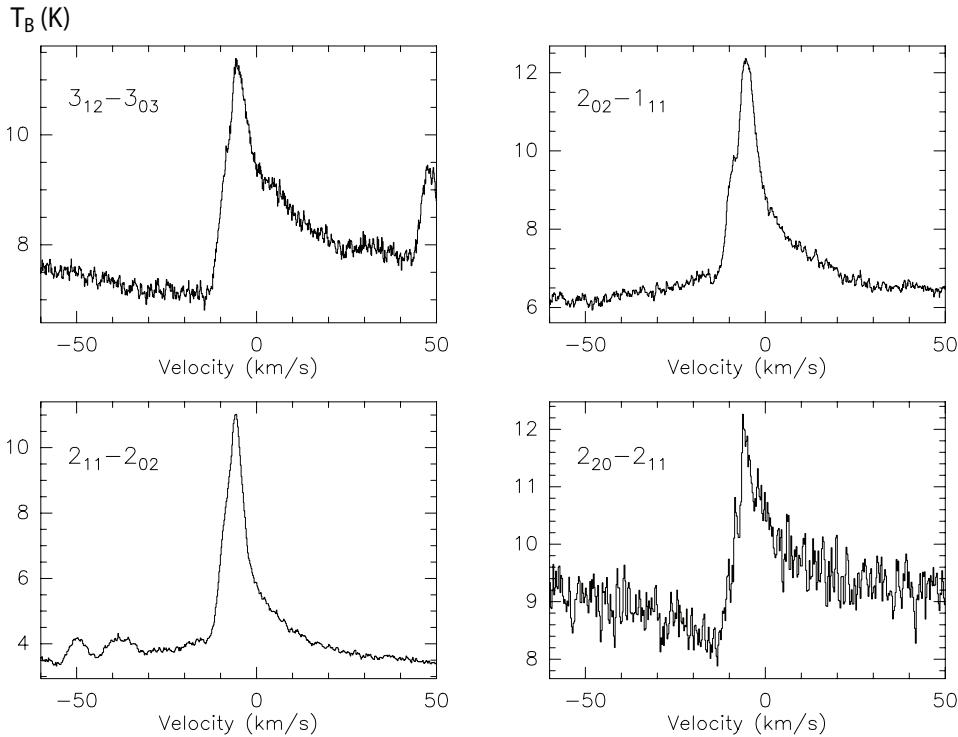}
\caption{Excited H$_2$O lines showing the narrow outflow component. The intensity scale is brightness temperature, in K, corrected for the aperture efficiency.} \label{NOspec}
\end{center}
\end{figure*}

In the $2_{20} - 2_{11}$ and $3_{12} - 3_{03}$ transitions, the warm and dense outflow shows a P-Cygni profile, but the blue wing is also weaker than the red wing in the remaining two transitions. The intensity difference between the red and blue wings is approximately 30\% of the continuum in all four transitions, leading to the conclusion that all four lines are optically thick ($\tau > 3$). The lack of any corresponding outflow features in the H$_2^{18}$O lines gives an upper limit of 80 for the optical depth.

The red lobe of the outflow, which is expected to reside behind the continuum source (the blue wing is seen in absorption, while the red one is not) can be used to determine the physical conditions in the outflowing gas. The integrated intensities emitted by the warm and dense outflow are listed in Table~4.

\begin{table}
\begin{center}
\caption{Integrated line intensities of H$_2^{16}$O transitions detected in the narrow outflow. \label{ofs}}
\begin{tabular}{lcc}
\hline\hline
$Transition$ & $Frequency$ & $\int T_{B}\,dv$  \\
& (GHz) & (K\,km\,s$^{-1}$)\\ 
\hline
$2_{02}-1_{11}$ & 987.9  & $24.1\pm2.4$\\
$2_{11}-2_{02}$ & 752.0  & $26.2\pm2.6$\\
$2_{20}-2_{11}$ & 1228.8 & $10.2\pm1.0$\\
$3_{12}-3_{03}$ & 1097.4 & $22.7\pm2.3$\\
\hline
\end{tabular}
\end{center}
\end{table}  

Using Radex (van der Tak et al. 2007), which calculates non-LTE radiative transfer, with the collisional rates from Faure et al. (2007), and assuming an isothermal, homogeneous medium, we derive the temperature and density in the outflow. Input parameters for the Radex models are: the background radiation temperature, the gas kinetic temperature, the H$_2$ density, the column density of water molecules, and the line width. For these parameters, the excitation temperature, optical depth, and radiation temperature of each line are calculated.

We set the background temperature to 2.73~K and the line width to 30 km\,s$^{-1}$, a value which we derive from a fit of several outflow features. Since we do not know the size, and thus the beam dilution of the outflow component, the observed line intensities do not provide good constraints. 
%Furthermore, all these lines are optically thick ($\tau \geq 3$), and thus the intensity ratios are not very sensitive to the column density. 
For this analysis we use an H$_2$O column density of $1 \times 10^{18}$ cm$^{-2}$, resulting in an optical depth of $\sim 25$, well within the limits $3 < \tau < 80$. A $\chi^2$ analysis of the model calculations reveals that the observations are best reproduced by an H$_2$ density of $2.5 \times 10^7$ cm$^{-3}$ and a kinetic temperature of 130~K. These results are in agreement with the non-detection of the dense outflow in the highly-excited lines. Whereas the H$_2$ density of the dense outflow component is quite well defined, with possible solutions ranging from $1.5 \times 10^7$ to $5.0 \times 10^7$ cm$^{-3}$, the temperature is less constrained and acceptable fits can be found for values between 60 K and 200~K.

The water column density in the dense outflow is not well constrained either, since all four H$_2^{16}$O lines are optically thick. Taking the limits for $\tau$ given above and the best fit results for the density and temperature, we derive a water column density between $5.0 \times 10^{16}$ and $1.2 \times 10^{18}$ cm$^{-2}$. Using the upper limit for the H$_2$O column density does not change the results for temperature and the H$_2$ density significantly. The best result for the lower limit on the H$_2$O column density can be found at a temperature of 94 K and an H$_2$ density of $3.6 \times 10^8$ cm$^{-2}$. Taking the uncertainty due to the unknown H$_2$O column density into account, we derive a temperature between 60 K and 200 K and an H$_2$ density higher than $2 \times 10^7$ cm$^{-2}$ for the dense outflow component.

\subsubsection{HDO in the broad outflow}

Besides H$_2^{16}$O, H$_2^{18}$O, and H$_2^{17}$O, which have been presented in paper I, the HDO $1_{11} - 0_{00}$ ground state transition is detected in absorption as well. In Figure~6 we show a comparison of the HDO and H$_2^{17}$O line profiles. The absorption feature covers the velocity range from --6 km\,s$^{-1}$ to approximately --50 km\,s$^{-1}$ and thus originates in the blue lobe of the broad outflow.

To calculate the column density of HDO in the outflow, we use equations (1) and (2), assuming that all molecules are in the ground state. This assumption is justified by the non-detection of emission in the red lobe of the outflow, which sets an upper limit for the HDO excitation temperature. 

The main difference between the calculation of the column density here and in foreground clouds (\S 3.2) is that the broad outflow does not cover the entire continuum source (see paper I). From the absorption features of H$_2^{16}$O and  H$_2^{18}$O $1_{10} - 0_{00}$, we determined that the H$_2^{16}$O $1_{10} - 0_{00}$ transition is optically thick, but the absorption feature lowers the intensity by only 17\%. Thus we concluded in paper I, that only 17\% of the continuum is covered by the outflow. To account for this low continuum coverage we have subtracted 83\% of the continuum from $T_{B}$ and $T_{back}$ in equation (1).

This calculation leads to an HDO column density of $1.17 \pm 0.4 \times 10^{13}$ cm$^{-2}$ and a para-H$_2^{17}$O column density of $1.21 \pm 0.4 \times 10^{13}$ cm$^{-2}$. The HDO/p-H$_2^{17}$O ratio is thus $1.0 \pm 0.5$. Assuming an ortho/para ratio of three (see paper I), and an H$_2^{16}$O/H$_2^{17}$O ratio of 1600, the resulting HDO/H$_2^{16}$O ratio is $2.1\pm 1 \times 10^{-4}$. This value is approximately an order of magnitude above the cosmic D/H ratio, but much lower than the HDO/H$_2$O ratios found in low-mass protostellar cores ($\sim 10^{-2}$; e.g., Codella et al. 2010).  

A source of significant uncertainty in the HDO/H$_2$O ratio is the assumption that all molecules are in their respective ground states. An higher excitation temperature would change the resulting HDO and H$_2^{17}$O column densities. However, the influence of the excitation temperature on the abundance ratio is limited. The upper state energies of the two transitions differ by only 25\%, and thus the relative population of the corresponding upper levels is very similar for any value of $T_{ex}$. Furthermore the moderate optical depth of both lines limits the uncertainty due to $J(T_{ex})$ in equation (1). Overall, we estimate that the uncertainty of the HDO/H$_2$O ratio, caused by the uncertainty in $T_{ex}$ is less than 20\%.

\begin{figure}
\begin{center}
\includegraphics[angle=0,width=0.9\columnwidth]{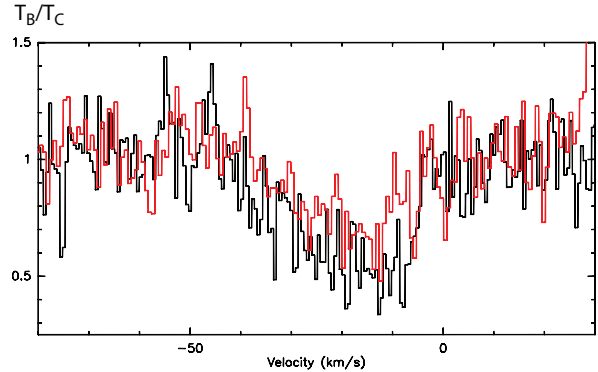}
\caption{Spectra of the $1_{11}-0_{00}$ line of H$_2^{17}$O (black) and HDO (red) toward NGC~6334~I, normalized by the continuum. \label{HDOout}}
\end{center}
\end{figure}

\subsection{Dense Cores}\label{DC}

Hot, dense cores in NGC 6334~I have been observed interferometrically (Hunter et al. 2006; Beuther et al. 2007, 2008) and their LSR velocities range from --7.6 to --8.1 km\,s$^{-1}$, which is 1.4 to 1.9 km\,s$^{-1}$ lower than the systemic velocity of the surrounding envelope. Therefore we use the velocity of the emission lines to identify the transitions originating from these dense cores.

\subsubsection{H$_2^{16}$O lines}

Five clearly detected and one tentatively detected H$_2^{16}$O lines are identified to stem from the dense cores. Line parameters of these transitions are listed in Table~5. All five lines have upper state energies of 450~K or higher, indicating the presence of hot gas. Furthermore, none of these lines show any indication of an outflow, or a second component, suggesting that only the dense cores contribute to the observed emission. Unfortunately all but two of these lines are known to show maser activity in certain interstellar environments (Maercker et al. 2008; Neufeld \& Melnick 1991). Therefore the local density and temperature structure, as well as the local radiation field, influence the observed line intensities. None of these transitions show particularly narrow lines, and thus no compelling sign of maser activity, yet no solution assuming LTE can be found. The fact that the $6_{24} - 7_{17}$ H$_2$O line, the line with the highest upper state energy, is much stronger than predicted by any LTE model with temperatures applicable to dense molecular clouds ($< 1000$ K) indicates that the excitation is clearly out of thermal equilibrium. We also attempted to model the line emission using Radex (\S 3.3.1), and with reasonable assumptions (kinetic temperature $< 1000$ K, $n$(H$_2$) from $10^5 $ to $10^8$ cm$^{-3}$) could not find a satisfactory solution for the hot core. The Radex models under-predict, similarly to the LTE models, the highest energy $6_{24} - 7_{17}$ transition by 1--2 orders of magnitude. Furthermore, for all lines except the $4_{22} - 4_{13}$ transition, the calculated level population is either inverted, or the upper level is more populated than expected for reasonable values of the gas kinetic temperature. All these findings suggest that a full radiative transfer model, including pumping by the dust continuum radiation is necessary to fully explain the H$_2^{16}$O emission from the hot core in NGC~6334~I (see \S 4).

\begin{table*}
\begin{center}
\caption{Line parameters of the H$_2^{16}$O lines associated with the dense core. \label{HC16}}
\begin{tabular}{lcccc}
\hline\hline
$Transition$ & $T_{B}$ & v$_{LSR}$ & $\Delta v$ & $E_{up}$ \\
& (K) & (km\,s$^{-1}$) & (km\,s$^{-1}$) & (K) \\ 
\hline
$4_{22}-4_{13}$ & $1.9\pm0.4$ & $-8.0\pm0.1$ & $5.4\pm0.3$ & 454.5\\
$4_{22}-3_{31}$ & $0.85\pm0.06$ & $-7.8\pm0.1$ &$8.1\pm0.2$& 454.5\\
$5_{24}-4_{31}$ & $0.7\pm0.1$ & $-8.3\pm0.3$ &$6.1\pm0.4$ & 599.1\\
$5_{32}-4_{41}$ & $0.52\pm0.02$ & $-8.6\pm0.6$ &$7.8\pm0.1$ & 732.4\\
$6_{24}-7_{17}$ & $0.09\pm0.02$  & $-8.0\pm0.2$ &$6.8\pm0.2$ & 867.7\\
$6_{34}-5_{41}^a$ & $0.43\pm0.27$  & $-6.9\pm0.5$ &$2.9\pm0.5$ & 934.2\\
\hline
\end{tabular}
\end{center}
$^a$Tentative detection.
\end{table*}

\subsubsection{Rarer Isotopologues}

The dense core is not seen in any ground state transition, because of the high optical depth of the surrounding envelope material. However, moderately excited lines ($J_u = 2$ or 3) of the rarer isotopologues H$_2^{18}$O and H$_2^{17}$O reveal the dense core emission. Contrary to the corresponding H$_2^{16}$O lines, these moderately excited transitions of the rarer isotopologues do not show any sign of outflows or multiple components and the central velocity of these lines confirms that they stem from the dense core (see Table~6).

\begin{table*}
\begin{center}
\caption{Line parameters of the rare water isotopologues in the dense core.\label{LineHc}}
\begin{tabular}{lcccc}
\hline\hline
$Line$ & $Frequency$ & $T_{B}$ & $v_{LSR}$ & $\Delta v$  \\
& (GHz) & (K) & (km\,s$^{-1}$) & (km\,s$^{-1}$)  \\ 
\hline
$\rm H_2^{18}O~2_{02}-1_{11}$ & 994.7 & $0.6\pm0.13$ & $-8.0\pm0.13$ & $4.7\pm0.26$ \\
$\rm H_2^{18}O~2_{11}-2_{02}$ & 745.8 & $0.47\pm0.05$ & $-7.95\pm0.08$ & $6.3\pm0.22$ \\
$\rm H_2^{18}O~3_{12}-3_{03}$ & 1095.6 & $0.6\pm0.11$ & $-8.8\pm0.12$ & $4.8\pm0.30$ \\
$\rm H_2^{17}O~2_{11}-2_{02}$ & 748.5 & $0.39\pm0.05$ & $-8.4\pm0.13$ & $6.1\pm0.44$ \\
$\rm H_2^{17}O~3_{12}-3_{03}$ & 1096.4 &$0.36\pm0.09$ & $-8.4\pm0.14$ & $4.1\pm0.31$ \\
$\rm HDO~2_{02}-1_{01}$ &919.3 & $0.25\pm0.06$ & $-8.2\pm0.17$ & $3.9\pm0.40$ \\
$\rm HDO~2_{11}-1_{10}$ & 1009.9 &$0.29\pm0.10$ & $-8.3\pm0.20$ & $4.6\pm0.46$ \\
$\rm HDO~2_{11}-2_{02}$ & 599.9 & $0.24\pm0.03$ & $-8.5\pm0.13$ & $6.1\pm0.34$ \\
\hline
\end{tabular}
\end{center}
\end{table*}

Figure~7 displays moderately-excited lines of H$_2^{16}$O, H$_2^{18}$O, and H$_2^{17}$O. Whereas the H$_2^{16}$O line is approximately a factor of ten stronger than the rarer isotopic lines, the corresponding H$_2^{18}$O and H$_2^{17}$O lines have comparable line strengths. The H$_2^{18}$O/H$_2^{17}$O ratios are 1.21 and 1.66 for the $2_{11} - 2_{02}$ line and the $3_{12} - 3_{03}$ line, respectively. This may be explained by an optically thick  dense core embedded in a larger, optically thin envelope, only seen in H$_2^{16}$O. Assuming similar excitation conditions for H$_2^{18}$O and H$_2^{17}$O, which is justified by the high densities expected in the dense core, and an H$_2^{18}$O/H$_2^{17}$O abundance ratio of $3.2 \pm 0.2$ (Wilson \& Rood 1994) leads to $\tau$(H$_2^{17}$O) = 1.78 and 0.81 for the $2_{11} - 2_{02}$ and $3_{12} - 3_{03}$ lines, respectively. In addition to H$_2^{18}$O and H$_2^{17}$O, lines of HDO have also been detected (see Table~6). The centroid velocity of the HDO lines (--8.2 km\,s$^{-1}$) confirms that these lines originate from the dense core, as well.

We analyze the observed lines of all three rare water isotopologues, H$_2^{18}$O, H$_2^{17}$O, and HDO independently, using XCLASS (Schilke et al. 1999; Comito et al. 2005), which calculates the level population assuming LTE, and from there the optical depth, $J$($T_{ex}$), and $T_{B}$. The input parameters are the source size, excitation temperature, column density, velocity, and line width.  
LTE is in general not a valid assumption for water lines, due to their high critical densities ($\sim 10^7$ cm$^{-3}$), and thus any LTE analysis of water lines has to be treated with caution. In case of a non-LTE level population, both the temperature and the column density derived may be largely underestimated. However, as shown later by full radiative transfer modeling (see \S~4) under the physical conditions present in NGC 6334~I, the water abundance is reproduced quite well assuming LTE.  

For our analysis we use a velocity of --8.2 km\,s$^{-1}$ and a line width of 4 km\,s$^{-1}$. The resulting size of the dense cores is $4 \pm 1^{\prime\prime}$, in good agreement with interferometric observations (Beuther et al. 2007; Hunter et al. 2006). The derived excitation temperatures and column densities are listed in Table~7: the best fit for H$_2^{17}$O and HDO is 50~K, while a slightly lower value of 43~K is derived for H$_2^{18}$O. The two estimates are consistent, given the uncertainties.  

Assuming an H$_2^{16}$O/H$_2^{18}$O ratio of 500, the XCLASS result leads to an H$_2$O column density of $1.5 \times 10^{18}$ cm$^{-2}$. With a source diameter of $4^{\prime\prime}$ and a distance of 1.7 kpc, we derive a total number of water molecules in the dense core of $1.22 \times 10^{52}$. Hunter et al. (2006) detected four dense cores within NGC 6334~I, of which two are hot ($T \geq 75$~K). Only these two cores may be emitting in the high energy water lines. The derived range of masses depends on the assumed dust temperature. Excluding scenarios cooler than 50~K, i.e., cooler than the excitation temperature of H$_2^{17}$O and HDO, the total mass of the hot cores is between 20 and 60 M$_\sun$, which yields a water abundance of $1.7 \times 10^{-6}$ with a factor of 2 accuracy. This water abundance is about two orders of magnitude below the value found in some other hot cores (e.g., Herpin et al. 2012). The measured abundance ratio of H$_2^{18}$O/H$_2^{17}$O is $3.75 \pm 1$, in good agreement with the $^{18}$O/$^{17}$O isotopic ratio (Wilson \& Rood  1994). HDO is a factor of 2 less abundant than H$_2^{17}$O, and therefore has abundance similar to that in the outflow, a factor of ten higher than the cosmic D/H ratio.

\begin{table}
\begin{center}
\caption{Excitation temperature and column density of H$_2^{18}$O, H$_2^{17}$O, and HDO derived with XCLASS.\label{xclasstab}}
\begin{tabular}{lcc}
\hline\hline
Molecule & $T_{ex}$ & $N$ \\
& (K) & (cm$^{-2}$) \\ 
\hline
H$_2^{18}$O & 43$\pm 3$ & $(3.0\pm0.7)\times 10^{15}$ \\
H$_2^{17}$O & 50$\pm 4$ & $(8.0\pm1.0)\times 10^{14}$ \\
HDO & 50$^{+20}_{-5}$ & $(4.0\pm1.0)\times 10^{14}$ \\
\hline
\end{tabular}
\end{center}
\end{table}

\begin{figure*}
\begin{center}
\includegraphics[angle=0,width=0.8\textwidth]{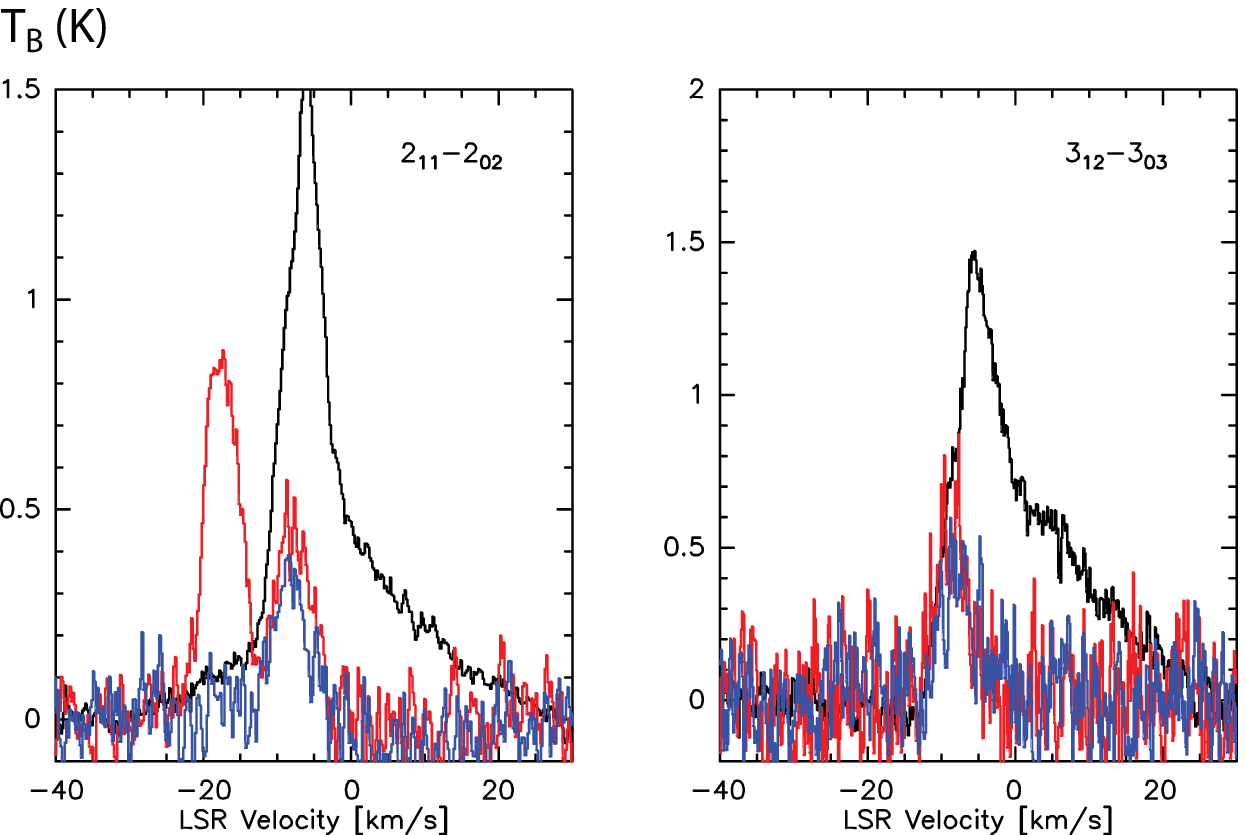}
\caption{(Left) Spectra of the $2_{11}-2_{02}$ line of H$_2^{16}$O (multiplied by 0.2, black), H$_2^{18}$O (red) and H$_2^{17}$O (blue). The line at --17 km\,s$^{-1}$ is emitted by dimethyl ether. (Right)  Spectra of the $3_{12}-3_{03}$ line of H$_2^{16}$O (multiplied by 0.5, black), H$_2^{18}$O (red) and H$_2^{17}$O (blue).  }\label{MR0}
\end{center}
\end{figure*}

\section{Radiative Transfer Models}\label{RT}

To model  H$_2^{16}$O lines, we use the spherically symmetric Monte Carlo radiative transfer code RATRAN (Hogerheijde \& Van der Tak 2000). The source model consists of a number of shells with varying H$_2$ density, water abundance, temperature, turbulent velocity, and expansion velocity. From such a source model, RATRAN calculates a grid of spectra integrated along a pencil beam, which has to be convolved to the required spatial resolution. The molecular data (Einstein coefficients and collisional rates calculated by Faure et al. 2007) are taken from the LAMDA database (Sch\"oier et al. 2005). The source structure is taken from Rolffs et al. (2011), based on APEX line and continuum observations, assuming a power-law density structure and solving for the temperature distribution self-consistently. This spherically symmetric model does not account for the outflow. Since the outflow features are dominant in several H$_2^{16}$O spectra, we had modified the source model accordingly.

Simply adding shells with a corresponding expansion velocity to the existing source model does not reproduce the outflow features appropriately. Such an approach fails, because, as shown in paper I and \S 3.3.2, only 17\% of the continuum is covered by the optically thick outflow. Furthermore, the outflow is collimated almost along the line of sight, whereas a modeling approach with an expanding shell would reproduce rather an isotropic outflow.  To model the collimated outflow, we thus calculated two RATRAN models. The first model does not include any outflow features. The second one includes an isotropic outflow represented by a number of expanding shells. Before convolving the grid of spectra with the beam size of the observations, we combined these two models to simulate the desired geometry.

The RATRAN model without an outflow was calculated using the source model from Rolffs et al. (2011), with the water abundance as the only parameter. For the model with an outflow, we use the same input parameters as for the first model, but add a number of expanding shells representing the outflow. Temperature, density, and water abundance in these additional shells are treated as free parameters. The output of both model calculations is a grid of pencil-beam spectra. We combine these two grids by taking spectra from the model with an outflow for points on the grid less than $3.5^{\prime\prime}$ from the center, whereas, we use the results from the model without an outflow for points further away. The radius of $3.5^{\prime\prime}$ was chosen to cover approximately 20\% of the continuum, in agreement with previous results. This combined grid was convolved with a Gaussian corresponding to the respective telescope beam size.

An alternative, simpler approach is to take the output of the radiative transfer model without an outflow and add the result of a fourth order polynomial fit of the outflow based on the observed spectra, also correcting for the  difference between the model and observed continuum. Fits were done using velocities $> -12$ km\,s$^{-1}$ (blue lobe) and $< +3$ km\,s$^{-1}$ (red lobe). This method gives better constraints on water abundance in the hot core and quiescent envelope, because uncertainties due to the emission from the outflow are minimized, but the model is not self-consistent.

Assuming a constant water abundance, the model fails to reproduce the observed line profiles, and we are forced to vary the abundance with temperature. For gas above 100~K, we derive a water abundance of $6 \times 10^{-7}$, whereas below 100~K the abundance drops to $3 \times 10^{-9}$. The H$_2$O abundance below 100~K agrees fairly well with theoretical models, as well as results of other investigations (e.g., Boonman et al. 2003; Chavarr\'{\i}a et al. 2010). Doty et al. (2002) predicts an H$_2$O abundance $\sim10^{-7}$ even at dust temperatures above 100 K due to the destruction by ions, such as C$^+$, H$^+$, H$_3^+$ , and He$^+$. However, their model does not include formation of water on grain surfaces, or continuous sublimation of ices infalling from the cold envelope toward the center of the protostar. Other models, such as Ceccarelli et al. (1996) and Rodgers \& Charnley (2003) suggest a water abundance of $10^{-4}$ in warm gas, which is approximately two orders of magnitude higher then our result. However, it is important to note that the temperature and the amount of hot gas might be overestimated in the RATRAN model due to the assumed geometry. The assumed spherical symmetry results in a lower escape probability of the cooling photons, than for other geometries. This lower escape probability may lead to an overestimation of the temperature in the inner part of the star-forming region. The water abundance in the outflow cannot be determined, since the outflow model is mainly defined by features due to the warm and dense outflow, for which no independent mass estimate exists. However, the total number of water molecules in the outflow is $1.2 \times 10^{51}$, which results in a total outflow mass of $4.7 \times 10^{-2}$ M$_\sun$, assuming an H$_2$O abundance of $4.3 \times 10^{-5}$ as found in the broad outflow (paper I). Thus the mass is only about 5\% of the mass of the broad outflow as determined by Leurini et al. (2006), which may explain why this outflow feature is not detected in CO. We also compared the model results with observations of H$_2^{18}$O and H$_2^{17}$O, using an H$_2^{16}$O/H$_2^{18}$O abundance ratio of 500 and an H$_2^{16}$O/H$_2^{17}$O ratio of 1600.

The best fit results for our model are show in Figures~8 and 9. The results of models, where the outflow was only fitted and not actually modeled (blue lines) show that the main component is reproduced quite well. One exception is ortho-H$_2$O $3_{12} - 2_{21}$, which is a factor of three stronger than predicted by our model. Another exception is the para-H$_2$O $6_{24} - 7_{17}$ transition, which is predicted to be very week (0.015 K), with the observed line approximately seven times stronger (0.11 K at a velocity of --7.7 km\,s$^{-1}$). One explanation could be that the $6_{24} - 7_{17}$ line shows maser activity. Alternatively this line could be also blended with other lines, especially considering the rather large line width (6.5 km\,s$^{-1}$) and taking into account the richness of the spectrum (see Fig. 8). However, we could not identify potentially interfering lines in the major molecular spectroscopy databases (JPL, Pickett et al. 1998\footnote{http://spec.jpl.nasa.gov}; CDMS, M\"uller et al. 2005\footnote{http://www.astro.uni-koeln.de/cdms}).  

For lines with very high optical depths, i.e., ortho-H$_2$O $1_{10} - 1_{01}$ or para-H$_2$O $1_{11} - 0_{00}$, the model calculations predict some emission in the line wings, which is not seen in the observed spectra. Modeling the outflow is difficult. With the exception of the ortho-H$_2$O $1_{10} - 1_{01}$, where we cannot find an acceptable model, there is a \emph{qualitative} agreement between the model and observed line shapes (emission, absorption, or a P-Cygni profile). However, there are significant \emph{quantitative} differences between detailed line shapes and line strengths predicted by the model and the observations, which indicate a limitation of the one-dimensional approach. The lines of the rarer isotopologues are not well reproduced (Figure~10). The strengths of the modeled H$_2^{18}$O lines are typically only 60\% of the observed ones, whereas modeled H$_2^{17}$O lines are much too weak. This indicates that the warm material might reside in small, dense clumps and thus the optical depths of the water lines may be underestimated by the models. This shows that a much more realistic three dimensional radiative transfer model is required to fit the water lines properly, which in turn requires detailed interferometric spectral maps of the source.

\begin{figure*}
\begin{center}
\includegraphics[angle=0,width=0.9\textwidth]{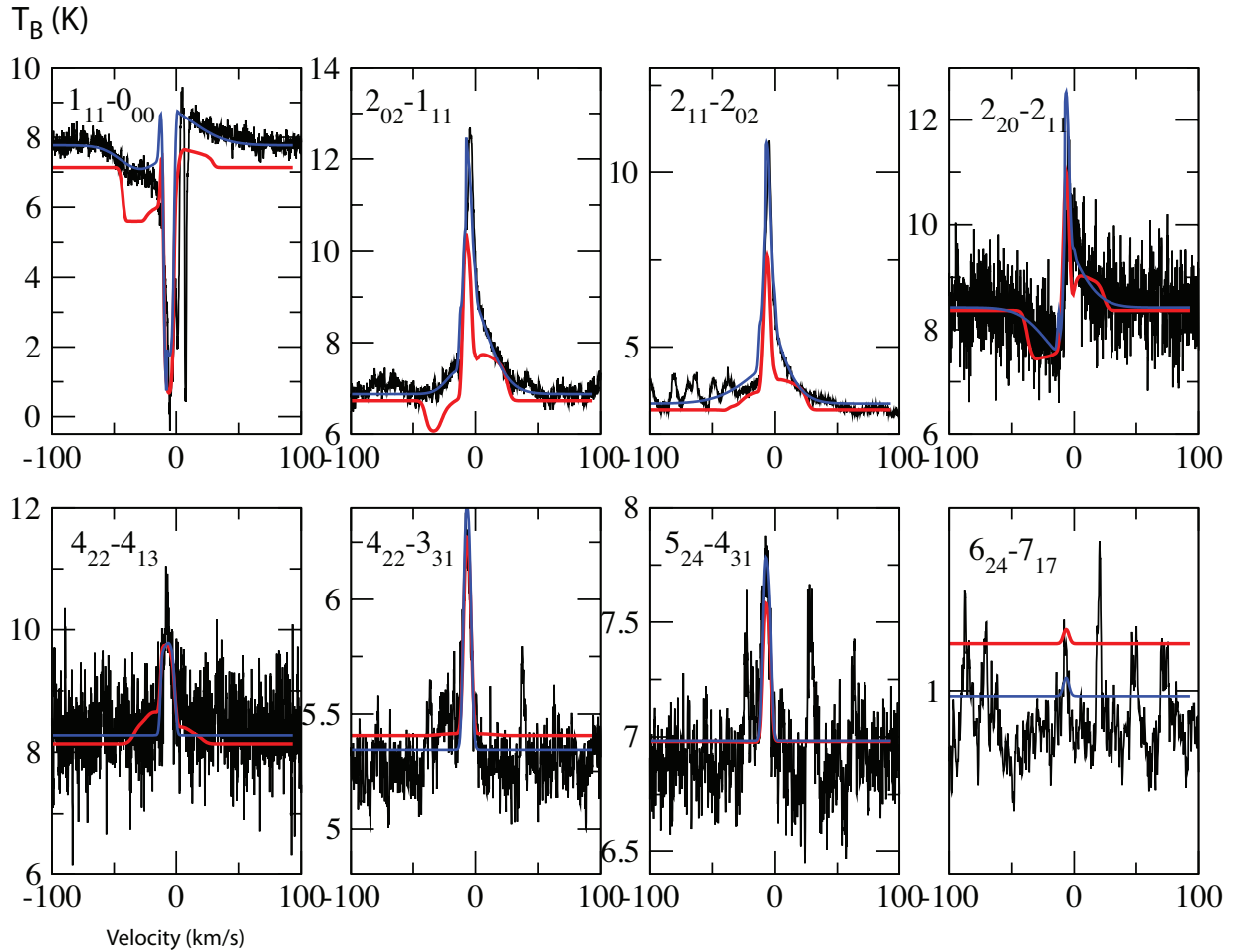}
\caption{Comparison of the radiative transfer model results with observations of para-H$_2^{16}$O (black). Blue lines represent models where the outflow component was fitted by a polynomial function, whereas the red lines show results for calculation  where the outflow was incorporated into the model. The intensity scale is brightness temperature, in K, corrected for the aperture efficiency. }\label{MR1}
\end{center}
\end{figure*}

\begin{figure*}
\begin{center}
\includegraphics[angle=0,width=0.9\textwidth]{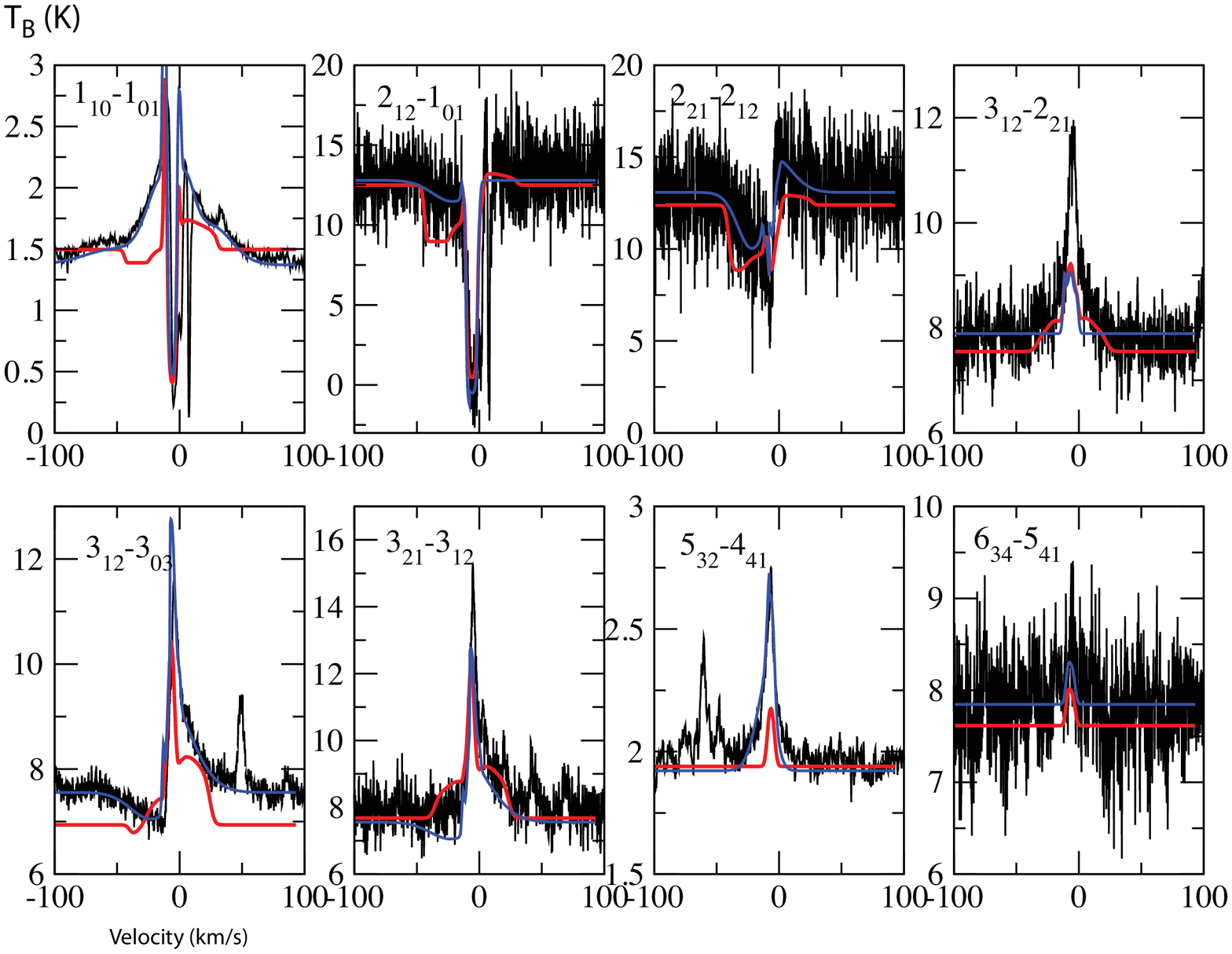}
\caption{Comparison of the radiative transfer model results with observations of ortho-H$_2^{16}$O (black). Blue lines represent models where the outflow component was fitted by a polynomial function, whereas the red lines show results for calculation  where the outflow was incorporated into the model. The intensity scale is brightness temperature, in K, corrected for the aperture efficiency.}\label{MR2}
\end{center}
\end{figure*}

\begin{figure*}
\begin{center}
\includegraphics[angle=0,width=0.9\textwidth]{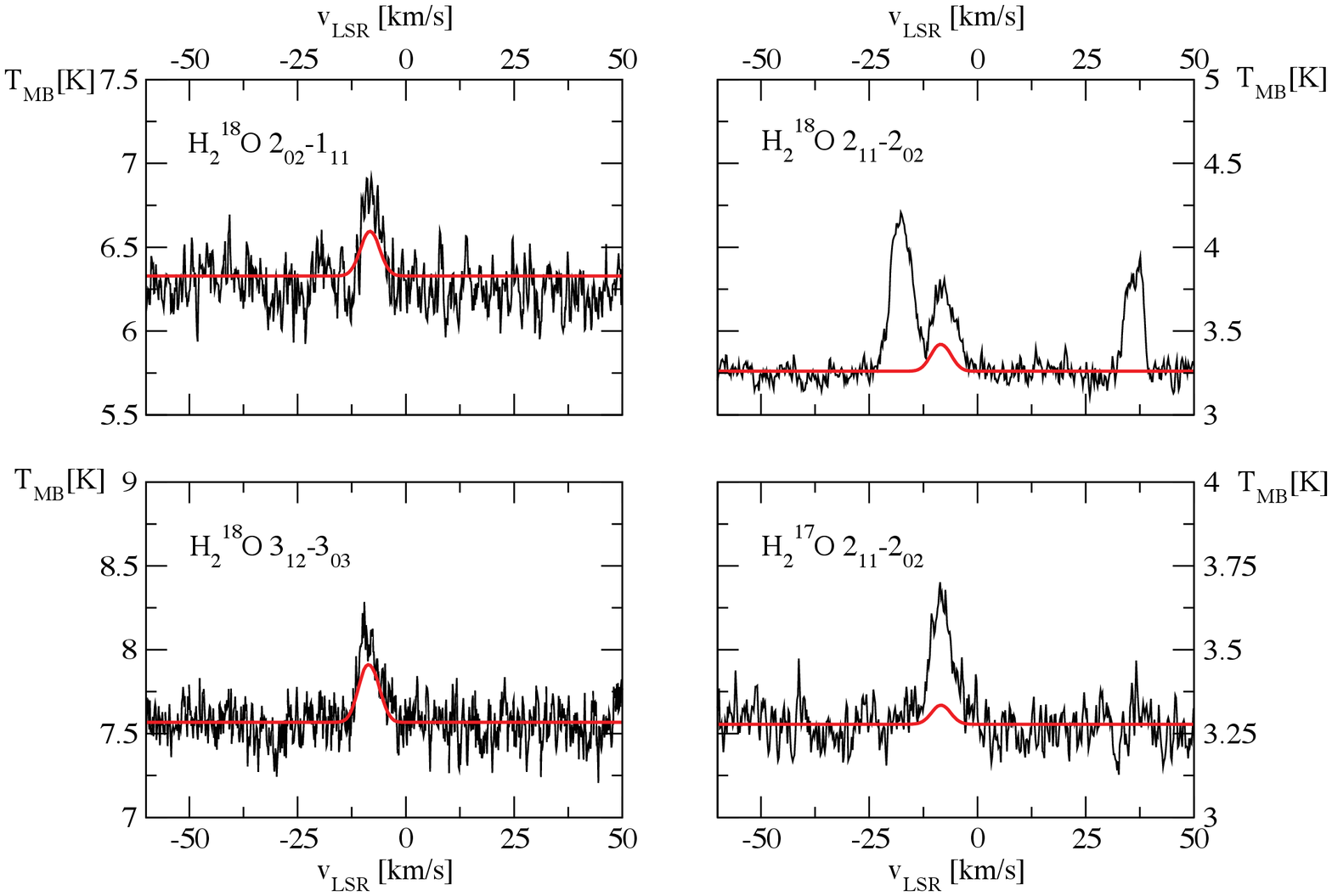}
\caption{Comparison of the radiative transfer model results (red) with observations (black) for H$_2^{18}$O and H$_2^{17}$O.}\label{MR3}
\end{center}
\end{figure*}

\section{Summary}\label{DIS}

The H$_2^{16}$O abundances in the cold envelope ($< 100$~K) and the foreground clouds of $3 \times 10^{-9}$ and $1.5 \times 10^{-8}$, respectively, agree fairly well with model predictions for cold regions (e.g., Ceccarelli et al. 1996, Doty et al. 2002), where freeze-out takes place. Chemical models also reproduce the water abundance of $4 \times 10^{-5}$ found in the broad outflow component, where we expect water ices to be sputtered from the dust grains. However, the water abundance we derive in the hot core ($\sim10^{-6}$) is two orders of magnitude lower than the value predicted by chemical models for hot, high-density material. The hot core H$_2$O abundance in NGC~6334~I derived here is in agreement with HIFI studies of hot corinos in low-mass star forming regions (e.g., Visser et al. 2012; Kristensen et al. 2012), but in contrast to some previous studies of water in high-mass star forming regions (Melnick et al. 2010; Chavarr\'{\i}a et al. 2010; Herpin et al. 2012) that found water abundances of order $10^{-4}$ at temperatures above 100~K. A high water abundance in the warm gas is only expected if all water molecules sublimate from the dust grains and stay as H$_2$O in the gas phase. A possible explanation for the low water abundances in NGC~6334~I are time dependent effects---water molecules may not have had enough time to fully desorb from the dust grains, or are quickly destroyed in the gas phase. Another possibility is that the dust temperature is moderate, near the threshold for the water ice sublimation, and therefore a significant fraction of water molecules may still reside on the dust grains. In addition, the presence of small-scale structures, such as disks or fragmented dense cores, may lead to overestimating the amount of the \emph{warm} H$_2$ based on dust continuum observations---if the H$_2$ column density is overestimated, the resulting H$_2$O abundance may be underestimated. Lines of the rare water isotopes suggest that the internal structure of this source may be more complex and, therefore, future, more detailed investigations of the H$_2$O abundance have to take into account the detailed structure of the embedded cores.

The two detected outflow components in NGC 6334~I are physically very different. The broad outflow, which has been previously detected in CO (Leurini et al. 2006), is of moderate temperature ($\sim 50$ K) and density ($n \sim10^4$ cm$^{-3}$). The narrow outflow ($T \sim 100$~K, $n \sim 2 \times 10^7$ cm$^{-3}$), which, to our knowledge, has not been identified in other tracers, is warmer and denser, which confirms that water is an excellent tracer of dense outflows (Santangelo et al. 2012). Whether these two components are actually two different outflows, or different regions within the same outflow cannot be conclusively answered. The fact that the central velocity of both outflow component is --10 km\,s$^{-1}$ points to the latter interpretation.

The ortho/para ratio in water is $3 \pm 0.5$ in the foreground clouds, as well as the outflow. This is consistent with the scenario, in which water molecules are formed with an ortho/para ratio of three, and the reactions which would cause a spin flip, such as those with H$^+$ and H$_3^+$, are of minor importance. The abundances of the rarer water isotopologues are consistent with the elemental abundance ratios. The H$_2^{18}$O/H$_2^{17}$O ratio in the dense core is $3.75 \pm 1$, in agreement with the ratio of 3.7 found in the broad outflow component (paper I) and, within the uncertainties, consistent with the $^{18}$O/$^{17}$O isotopic ratio.

HDO is detected in the hot core, as well as the broad outflow component. The HDO/H$_2$O ratio in both regions is $\sim 2 \times 10^{-4}$, a factor ten above the cosmic abundance of deuterium, and close to the value found in Earth oceans. However, this ratio is much lower than that typically found in low-mass star forming regions ($\sim 0.04$, Codella et al. 2010), and at the lower end of values found in other high-mass star forming regions (Bergin et al. 2010; Comito et al. 2003, 2010; Van der Tak et al. 2006). HDO is formed in excess in cold clouds, either in the gas-phase or on dust grains, where many deuterated molecules are abundant, whereas at higher temperatures ($\geq 50$~K) HDO/H$_2$O ratios similar to the one found here are predicted (Roberts \& Millar 2000). The high HDO abundance in Orion KL was interpreted as material recently desorbed from dust grains, and thus the high HDO/H$_2$O ratio reflects the physical conditions from the past. Thus the low HDO/H$_2$O ratio in NGC~6334~I, in agreement with the relatively low water abundance, is another indication that most water, especially water formed during the cold early stages of star-formation, is still frozen-out onto dust grains.

\acknowledgements

HIFI has been designed and built by a consortium of institutes and university departments from across Europe, Canada and the United States under the leadership of SRON Netherlands Institute for Space Research, Groningen, The Netherlands and with major contributions from Germany, France and the US. Consortium members are: Canada: CSA, U.Waterloo; France: CESR, LAB, LERMA, IRAM; Germany: KOSMA, MPIfR, MPS; Ireland, NUI Maynooth; Italy: ASI, IFSI-INAF, Osservatorio Astrofisico di Arcetri-INAF; Netherlands: SRON, TUD; Poland: CAMK, CBK; Spain: Observatorio Astronmico Nacional (IGN), Centro de Astrobiologa (CSIC-INTA). Sweden: Chalmers University of Technology MC2, RSS \& GARD; Onsala Space Observatory; Swedish National Space Board, Stockholm University Stockholm Observatory; Switzerland: ETH Zurich, FHNW; USA: Caltech, JPL, NHSC. Support for this work was provided by NASA through an award issued by JPL/Caltech, by l'Agence Nationale pour la Recherche (ANR), France (project FORCOMS, contracts ANR-08-BLAN-022), and the Centre National d'Etudes Spatiales (CNES). We thank the anonymous referee for constructive comments that have significantly improved the presentation.
%\bibliographystyle{apj.bst}
%\bibliography{bibtex/library.bib}
%\end{document}

\end{document}